\documentstyle [12pt]{article}

\begin{document}

\begin{center}

{\large \bf Gravitational Particle Production in Spinning Cosmic 
String Spacetimes} \\

\vspace{2cm}

V. A. De Lorenci ${}^{(a)}$, R. De Paola ${}^{(a),(b)}$ 
and N. F. Svaiter ${}^{(a)}$ \\
\vspace{1cm}
${}^{(a)}$ {\it Centro Brasileiro de Pesquisas F\'{\i}sicas,} \\
\vspace{0.1cm}
{\it Rua Dr. Xavier Sigaud 150, Urca,} \\
\vspace{0.1cm}
{\it CEP 22290-180 -- Rio de Janeiro -- RJ, Brazil.} \\
\vspace{0.7cm}
${}^{(b)}$ {\it Pontif\'{\i}cia Universidade Cat\'olica do Rio de Janeiro,} \\
\vspace{0.1cm}
{\it Rua Marqu\^es de S\~ao Vicente, 110, G\'avea} \\
\vspace{0.1cm}
{\it CEP 22453-000 -- Rio de Janeiro -- RJ, Brazil.} \\

\vspace{2cm}

\begin{abstract}

The spontaneous loss of angular momentum of a spinning cosmic string due to 
particle emission is discussed. The rate of particle production between 
two assymptotic spacetimes: the spinning cosmic string spacetime in the 
infinite past and a non-spinning cosmic string spacetime in the infinite  
future is calculated.

\end{abstract}

\end{center}

\vspace{1cm}

Pacs numbers: 03.60.Bz, 04.20.Cv, 11.10.Qr

\newpage

There is a broad class of predictions of various models in grand 
unified theories (GUT\rq s). Between them, phase transitions may 
result in the formation of cosmic strings \cite{Kibble}. They are 
extended one dimensional infinite topological defects. The cosmic 
strings affect the spacetime mainly topologically, given a 
conical structure to the space region around the cone of the string. 
The conical topology  may be responsible for several gravitational effects, e.g.,  
gravitational lense \cite{Vilenkin} and particle
production due to the changing gravitational field during the formation 
of such object \cite{Parker}.
There are a lot of papers studying quantum processes in a cosmic 
string spacetime. Of special interest for us are the following: Ref. \cite{Harari} 
where pair production in a straight cosmic string spacetime is discussed, 
Ref. \cite{Nami} where the rate of transition of a two-level system coupled 
with a scalar field in the presence of a cosmic string is analysed,
and finally Ref. \cite{Mazur} where spinning cosmic string spacetime 
is investigated.

In this paper we are interested in evaluate the particle production due
to the changing of the gravitational field in the situation of gradual loss
of angular momentum of a spinning cosmic string.  
We set $\hbar=k_{B}=c=1$.

Let us consider the following metric structure for the spacetime in 
the exterior  of a straight cosmic string:
\begin{equation}
{\rm d}s^{2} = 
\left\{ {\rm d}t + \zeta(t){\rm d}\varphi \right\}^{2}-{\rm d}r^{2}
- b^{2}r^{2}{\rm d}\varphi^{2}-{\rm d}z^{2},
\label{V1}
\end {equation}
where $x^{\mu}=\{t,r,\varphi,z\}$ are the usual cylindrical coordinates
with the range: $(0\leq r<\infty,\, -\infty<z<\infty,\, 
0 \leq \varphi \leq 2\pi)$. 
The constant $b$ is called the conical parameter and is related with
the deficit angle of the conical singularity by
\begin{equation}
b=1-4\mu G, 
\label{2}
\end{equation}
where $\mu$ is the linear density of the string.
The function $\zeta(t)$ appearing in Eq. (\ref{V1}) is defined by
\begin{equation}
\zeta(t) \equiv 2GJ\left[1-\tanh\left(\frac{t}{t_{0}}\right)\right],
\label{V2}
\end{equation}
and the others constants are the gravitational Newton constant ($G$) and
the angular momentum ($J$) of the source.

From relation (\ref{V2}) we can stablish the following asymptotics 
conditions for the metric structure:
\begin{eqnarray}
\lim_{t\rightarrow +\infty}\zeta(t) &=& 4GJ,
\label{V4} \\
\lim_{t\rightarrow -\infty}\zeta(t) &=& 0.
\label{V5}
\end{eqnarray}
In the two asymptotic regions --- the infinite past and infinite future ---
the spacetime metric structure reduces to:
\begin{eqnarray}
{\rm d}s^{2}_{-\infty} &=& 
\left( {\rm d}t + 4GJ{\rm d}\varphi\right)^{2}-{\rm d}r^{2}
- b^{2}r^{2}{\rm d}\varphi^{2}-{\rm d}z^{2},
\label{V6}\\
{\rm d}s^{2}_{+\infty} &=& 
{\rm d}t^{2}-{\rm d}r^{2}-b^{2}r^{2}{\rm d}\varphi^{2}-{\rm d}z^{2}.
\label{V7}
\end{eqnarray}
As it is well know, the above metrics represent the structure of the
spacetime in the exterior region of a rotating and a non-rotating
cosmic string\footnote{
By transforming the azimutal coordinate $\theta=b\varphi$ the line 
element reduces to $${\rm d}s^{2}={\rm d}t^{2}-{\rm d}r^{2}-
r^{2}{\rm d}\theta^{2}-{\rm d}z^{2},$$
where the azimutal angle is defined in the interval $0\leq \theta\leq 2\pi b$.
}, respectively. 

With this picture in mind we will analyse the rate of particle produced by 
changing the gravitational field between the two asymptotic spacetimes.
The same idea it was used in a toy model by Bernard and 
Duncan \cite{Duncan}. These authors studied a two dimensional 
Robertson-Walker model where the
conformal scale factor has the same functional form as Eq. (\ref{V2}). 
In the two asymptotics limits, the spacetime becomes Minkowskian, and 
these authors were able to obtain the modes solutions of the Klein-Gordon
equation in these two limits. A straighforward calculation of the Bogoliubov 
coefficients between the {\bf in} and {\bf out} modes gives the rate of
particle production during the expansion of the universe.

In this paper we will develop a similar idea.
The mathematical treatment will follows the same lines a Bernard and Duncan 
paper.
We consider the case of a massive minimally coupled Hermitian scalar 
field $\phi(t,\vec{x})$ defined at all points of the $4$-dimensional 
spacetime with line element given by eq.(\ref{V1}). 
The Klein-Gordon equation is given by:
\begin{equation}
\left[g^{\mu\nu}D_{\mu}D_{\nu} + M^{2}\right]\phi(t,\vec{x}) = 0,
\label{V8}
\end{equation}
where the symbol $D_{\alpha}$ represents the covariant derivative 
with respect to the metric $g_{\alpha\beta}$, and $M$ is the mass
of the quanta of the scalar field. For further reference we point out that
the determinant of the metric $g_{\alpha\beta}$ for the general
spacetime given by eq.(\ref{V1}) is: 
\begin{equation}
\det [g_{\mu\nu}]\equiv g = -b^{2}r^{2}.
\end{equation}

Here we are interested in studying the process of creation of particles
and radiation by the gravitational field changing during the evolution
of a cosmic string that looses angular momentum during the time.  
To mantain the particle produced in a limited region of the
space we impose Dirichlet boundary conditions at $r=R$,
\begin{equation}
\left.\phi(t,r,\varphi,z)\right|_{r=R} = 0,
\label{R}
\end{equation}
and periodic boundary conditions in $z$ with period $L$.

It is well know that the geometry given by Eq. (\ref{V6}) generate
closed time-like curves (CTC). 
In order to circumvent this problem, we impose an additional
vanishing boundary condition at $r=R_{0} > 4GJ/b$. Thus, the radial coordinate
has the domain $R_{0} < r < R$. For a carefull study how to construct
quantum field theory in a spacetime with CTC, see for instance
Ref.\cite{Boul}. 
The same problem appear in $(2+1)$ dimensional gravity since
the spinning cosmic string spacetime is exactly the solution of $(2+1)$ Einstein
equations of a spinning point source \cite{Deser}.
Deser, Jackiw and 't Hooft derived the solution to the $D=3$ Einstein 
gravity with a massless spining source. The generalization for 
massive spining sources was obtained by Clement \cite{Clement}. As 
we point out the solution show that in both cases the 
three dimensional geometry is the Minkowski space with a edge removed.
In this case a non-trivial physical situation arrises. The points that 
we have to identify across the deleted edge differ in the time coordinate 
by an amount proportional to the angular momentum of the source.

In the asymptotic past --- that corresponds a rotating cosmic string spacetime ---
the Klein-Gordon equation given by Eq. (\ref{V6}) reduces
to the form:
\begin{equation}
\left[\left(1-\frac{a^{2}}{b^{2}r^{2}}\right)\frac{\partial^2}{\partial t^2}
-\frac{1}{r}\frac{\partial}{\partial r}\left(r\frac{\partial}{\partial r}
\right)
-\frac{1}{b^{2}r^{2}}\frac{\partial^{2}}{\partial\varphi^{2}}
-\frac{\partial^{2}}{\partial z^{2}}
+ \frac{2a}{b^2r^2}\frac{\partial^2}{\partial t \partial \varphi}  
+M^{2}\right]\phi(t,r,\varphi,z)=0. 
\label{V9}
\end{equation}.

It is not difficult to find the mode solutions $u_{j}$ 
and they are given by:
\begin{equation}
u_{j}(t,\vec{x}) = N_{1}e^{-i\omega_{l}t}e^{ikz}e^{im\varphi}
J_{\mu}(qr),
\label{V16}
\end{equation}
with
\begin{eqnarray}
\mu &\equiv& \frac{|m+4GJ\omega_{l}|}{b},\\
q &=& \sqrt{\omega_{l}^{2} - k^{2} - M^{2}}
\label{V34}
\end{eqnarray}
and 
\begin{equation}
k = \frac{2\pi n}{L}.
\end{equation}
We choose the constant $N_{1}$ to make the set orthonormal. Thus,
\begin{equation}
N_{1} = (2\omega_{l})^{-\frac{1}{2}}\left\{ V\left[ J^{'}_{\mu}(qR) \right]^2
-  V_{0}\left[J^{'}_{\mu}(qR_{0})\right]^2 \right\}^{-\frac{1}{2}},
\label{V33}
\end{equation}
where we defined the 3-volumes $V \equiv b\pi LR^2$ and $V_{0} \equiv 
b\pi LR_{0}^2$. The values of $\omega_{l}$ are determined by the 
vanishing boundary conditions.

The modes $u_{j}(t,\vec{x})$ form a basis in the space of solutions of the 
Klein-Gordon equation and can be used to expand the field operator in the
following way:
\begin{equation}
\phi_{in}(\vec{x},t) = \sum_{j}\left\{ {\bf a}_{j} u_{j}(t,\vec{x}) 
+ {\bf a}_{j}^{\dag}u_{j}^{*}(t,\vec{x})\right\},
\label{V35}
\end{equation}
where we are using a collective index $j\equiv\{l,m,n\}$.

The creation and anihilation operators ${\bf a}^{\dag}_{j}$ and ${\bf a}_{j}$ 
satisfies the usual comutation relations:
\begin{equation}
[ {\bf a}_{j},{\bf a}^{\dag}_{j}] = \delta_{j,j^{'}},
\label{aa}
\end{equation}
and the in-vacuum state is defined by
\begin{equation}
{\bf a}_{j}|0,in> = 0 \,\,\,\, \forall\,\, j.
\label{0in}
\end{equation}

We can follow the same lines to canonical quantize the field in the 
infinite future. The Klein-Gordon equation in the non-rotating cosmic string 
spacetime given by eq.(\ref{V7}) reads
\begin{equation}
\left[\frac{\partial^2}{\partial t^2} -
\frac{1}{r}\frac{\partial}{\partial r}\left(r\frac{\partial}{\partial r}\right)
-\frac{1}{b^{2}r^{2}}\frac{\partial^{2}}{\partial\varphi^{2}}-
\frac{\partial^{2}}{\partial z^{2}}
+M^{2}\right]\phi(t,r,\varphi,z)=0.
\label{V13}
\end{equation}
The out modes solutions of the Klein-Gordon equation also form a 
complete set and can be used to expand the field operator as: 
\begin{equation}
\phi_{out}(t,\vec{x}) = \sum_{j}\left\{ {\bf b}_{j} v_{j}(\vec{x},t) 
+ {\bf b}_{j}^{\dag}v_{j}^{*}(t,\vec{x})\right\},
\label{V36}
\end{equation}
where the modes $v_{j}$ is defined by:
\begin{equation}
v_{j}(t,\vec{x}) = N_{2}e^{-i\Omega_{l}t}e^{ikz}e^{im\varphi}
J_{\nu}(\bar{q}r),
\label{V37}
\end{equation}
with
\begin{eqnarray}
\nu &\equiv& \frac{|m|}{b},\\
\bar{q} &=& \sqrt{\Omega_{l}^{2} - k^{2} - M^{2}}.
\label{nu}
\end{eqnarray}
Choosing the constant $N_{2}$ in order to make the set of modes $\{v_{j},v^{*}_{j}\}$ orthonormal, results:
\begin{equation}
N_{2} = (2\Omega_{l})^{-\frac{1}{2}}\left\{ V\left[ J^{'}_{\nu}(qR) \right]^2
-  V_{0}\left[J^{'}_{\nu}(qR_{0})\right]^2 \right\}^{-\frac{1}{2}}.
\label{V33}
\end{equation}

Similarly the creation and anihilation operators ${\bf b}^{\dag}_{j}$ and 
${\bf b}_{j}$ satisfies the usual comutation relation:
\begin{equation}
[ {\bf b}_{j},{\bf b}^{\dag}_{j}] = \delta_{j,j^{'}},
\label{bb}
\end{equation}
and the vacuum state in the out-spacetime, is defined by
\begin{equation}
{\bf b}_{j}|0,out> = 0 \,\,\,\, \forall \,\, j.
\label{0out}
\end{equation}

Following Parker we will calculate the rate of particle 
production between 
two asymptotic spacetimes discussed above: the spinning 
cosmic string spacetime in the infinite past and a unspinning cosmic string 
spacetime in the infinite future. 

A important point is that in our model we have not to due with
the problems of junction conditions since there is no sudden approximation
here. The metric evolves continuously between both asymptotics states. The
angular momentum of the spinning cosmic string is lost by 
particle emission processes. The fundamental quantity we have to calculate
is the Bogoliubov coefficients between the modes in the non-rotating and
rotating cosmic string spacetime. The average number of in-particles in the modes
$(l,m,n)$ produced by this process is given by:
\begin{equation}
<in,0| b^{\dag}_{j}b_{j} |0,in> = \sum_{i}\left|\beta_{ij}\right|^{2}.
\label{number}
\end{equation}
Using the definition of the Bogoliubov coefficients $\beta_{ij}$ given by
\begin{equation}
\beta_{jj'} = -(u_{j},v^{\dag}_{j'})
\end{equation}
we have 
\begin{equation}
\beta_{jj'} = -2\pi b L(\Omega_{l} +
 \omega_{l'})N_{1}N_{2}\xi(R,R_{0})\delta_{m,m'}\delta_{n,n'}
\label{beta}
\end{equation}
where
\begin{equation}
\xi(R,R_{0}) \equiv \int^{R}_{R_{0}} {\rm d}r\, r\,
 J_{\mu}(qr)\,J_{\nu}(\bar{q}r).
\end{equation}
Substituting (\ref{beta}) in Eq. (\ref{number}) and use the definitions of the
normalization constants $N_1$, $N_2$, the average number  of particles
in the modes $(l,m,n)$ produced is:
\begin{eqnarray}
<in,0| b^{\dag}_{j}b_{j} |0,in> &=& \pi b^2 L^2 \left[\left(\frac{\Omega_{l}}{\omega_{l'}}\right)^{\frac{1}{2}}
+ [\left(\frac{\omega_{l}}{\Omega_{l'}}\right)^{\frac{1}{2}}\right]^2 
\nonumber\\
&\times& \left\{ \left[VJ^{'}_{\mu}(qR)^2
+ V_{0}J_{\mu}^{'}(qR_{0})^2\right] \left[VJ^{'}_{\nu}(\bar{q}R)^2
+ V_{0}J_{\nu}^{'}(\bar{q}R_{0})^2\right] \right\}^{-1}
\end{eqnarray}

Let us sumarize the results obtained in the paper. We discussed 
particle production by lost of angular momentum in a spinning cosmic string
spacetime. To circumvent the problem of CTC\rq s we assume a cosmic string 
with a radius fixed. Moreover, to avoid the problem of square of distribution
we following Parker\rq s arguments impossing vanishing boundary conditions
a cylinder with finite radius $R$.

A possible continuation of this paper is to formulate the energy 
conservation law, that is to show if there is a balance between the 
total energy of the particles creation and the energy associated 
with loss of angular momentum. This can be done comparing the vacuum 
stress-tensor of the massive field in the spinning and non-spinning 
cosmic string spacetime. The calculation for a massless conformally coupled 
scalar field in the non-spinning cosmic string spacetime
has been done by many authors \cite{Linet}. The same calculation in 
the spinning cosmic string spacetime has been done by Matsas 
\cite{Matsas}. As far as we know the renormalized 
stress tensor of a massive minimally coupled scalar 
field  has not been investigated in the literature. The evaluation of such
quantite is fundamental to investigate the model taken into acount the back
reation problem. Note that the source of Einstein\rq s equations that generates
the line element in the non asymptotic limit in unknown. 
Actually, the particle production  must be included in the energy momentum 
tensor of the source and together with the matter energy momentum tensor satisfy the semi-classical Eintein\rq s 
equations. The calculation of the renormalized energy momentum tensor of the
massive scalar field and the corresponding energy balance in under investigation.

\section{Acknowledgement}

We would like to thank Dr.B.F.Svaiter for valuable comments. This 
paper was supported by Conselho Nacional de Desenvolvimento 
Cient\'{\i}fico e Tecnol\'ogico (CNPq) of Brazil.

\end{document}